\documentclass[11pt]{article}
\usepackage{amsfonts}
\usepackage{amsmath, amsthm, amssymb}
\usepackage{latexsym}
\usepackage{graphicx}
\usepackage{amscd}
\usepackage{pictexwd,dcpic}
\newtheorem{definition}{Definition}[section]

\begin{document} 

\title{Quantized Poker}
\author{Steven A. Bleiler\footnote{Department of Mathematics and Statistics, Portland State University, PO Box 751, Portland, Oregon 97207-0751 USA. Email: bleilers@pdx.edu.}}
\maketitle

\begin{abstract}
Poker has become a popular pastime all over the world.  At any given moment one can find tens, if not hundreds, of thousands of players playing poker via their computers on the major on-line gaming sites.  Indeed, according to the Vancouver, B.C. based pokerpulse.com estimates, more than 190 million US dollars daily is bet in on-line poker rooms. But communication and computation are changing as the relentless application of Moore's Law brings computation and information into the quantum realm. The quantum theory of games concerns the behavior of classical games when played in the coming quantum computing environment or when played with quantum information.  In almost all cases, the "quantized" versions of these games afford many new strategic options to the players. 

The study of so-called quantum games is quite new, arising from a seminal paper of D. Meyer \cite{Meyer} published in Physics Review Letters in 1999. The ensuing near decade has seen an explosion of contributions and controversy over what exactly a quantized game really is and if there is indeed anything new for game theory. With the settling of some of these controversies \cite{Bleiler}, it is now possible to fully analyze some of the basic endgame models from the game theory of Poker and predict with confidence just how the optimal play of Poker will change when played in the coming quantum computation environment.  The analysis here shows that for certain players, "entangled" poker will allow results that outperform those available to players "in real life".  
\end{abstract}
\section{ Overview}

When playing on line, players indicate their strategic choices, i.e. their choice of action at each of their opportunities to act, by communicating this choice of action to a central server that acts as a referee.  At the conclusion of each hand this server determines the outcome of the hand and makes the appropriate payoffs.  Of fundamental importance here will be how the players communicate their strategic choices to the central server.  

For simplicity, the models of a poker endgame we consider will allow each player a choice of exactly two strategic options.  These options will be represented as a state of a "bit", either classical or quantum as appropriate to our discussion.  The communication of strategies will be modeled as follows.  The referee prepares the players' bits in a particular initial state and sends them to the players.  The players then indicate their strategic choice by acting on the bit, the two pure strategies represented by the two actions on the bits of either "not flipping" the bit or of "flipping" it.  Then the players return their bits to the referee.  The referee then examines the bits, from these observations determines each player's strategic choice, and from these computes the outcome of the game and the appropriate payoffs.  As we shall shortly see, whether this communication occurs over classical or quantum channels will have a significant effect on the number and type of players' strategies and on the determination of the Nash equilibrium strategies (also known as "solutions" or "optimal strategies") for our models.

The endgame models considered here are known respectively as "Simplified Poker" and the "Nash-Shapley Poker Model".  Simplified Poker is  a discretized version of the poker models of Borel \cite{Borel} and Von Neuman \cite{VN}, and is a two-player game that illustrates the advantage of "mixing" one's strategic choices to achieve a maximal return via an "optimal bluffing frequency".  The Nash-Shapley poker model \cite{NS} is a three-player game that as well as further illustrating the advantage to "mixing" one's strategic choices, illustrates two other important points.  First, the Nash Shapley model illustrates advantage of what the poker literature terms "position".  In particular, the last player to act can exploit the informational advantage of acting last by bluffing at the appropriate frequency.  Second, and more importantly, the Nash-Shapley model also illustrates the importance of defending against the bluff, which represents strength when one is in fact weak, via \emph{slow-playing}, which represents weakness when one is in fact strong, via an "optimal slow-playing frequency".  The Nash-Shapley model also has the interesting feature in its optimal bluffing and slow-playing frequencies are irrational, and hence can only be approximated and never actually realized in real life play.

Our endgame models can be thought of as abstracting the following situation from Texas Hold'em.  After the last community card has been dealt, the common cards form a king high straight with no possible flush. Because of these cards the players cards are now one of exactly two types, either "High" (an ace, which makes an ace high straight the best possible hand given the community cards) or  "Low" (everything else, leaving the player with the king high straight on the board).  We further assume that from long experience playing against each other the players know that the probability that a given opponent holds an ace is exactly $\frac{1}{2}$. We'll consider the betting up to this point as an ante a, and that each player has the ability to make a single bet of size b.

This situation is abstracted as follows.  Each player antes an amount $a$ into a "pot" and is dealt a single card from an infinite deck containing two types of cards, $H$ and $L$.  For each player the probability of receiving an $H$ is exactly $\frac{1}{2}$, as is the probability of receiving an $L$.  In a showdown, $H$ beats $L$ and the winning hand receiving the pot, and if multiple players hold equal winning hands, the pot is divided equally between them. 

\section{Some Classical Game Theory}
Here it's best to begin with a definition.
\begin{definition}
Given a set $\{ 1, 2, \cdots, n \}$ of players, for each player a set $S_i$ $(i=1, \cdots, n)$ of so-called \emph{pure strategies}, and a set $\Omega_i$ $(i=1, \cdots, n)$ of \emph{possible outcomes}, a \emph{game} $G$ is a vector-valued function whose domain is the Cartesian product of the $S_i$'s and whose range is the Cartesian product of the $\Omega_i$'s. In symbols 
                       \[\prod_{i=1}^n S_i \stackrel{G}{\longrightarrow} \prod_{i=1}^n \Omega_i \]
The function $G$ is sometimes referred to as the \emph{payoff function}.
\end{definition}

Here a \emph{play} of the game is a choice by each player of a particular strategy $s_i$ the collection of which forms a \emph{strategy profile} $(s_1, \cdots, s_n)$ whose corresponding \emph{outcome profile} is $G(s_1, \cdots, s_n)=(\omega_1, \cdots, \omega_n)$, where the $\omega_i$'s represent each player's individual outcome. Note that by assigning a real valued \emph{utility} to each player which quantifies that player's preferences over the various outcomes, we can without loss of generality, assume that the $\Omega_i$'s are all copies of $\mathbb{R}$, the field of real numbers.

In game theory one is frequently concerned with the identification of special strategies or strategic profiles. For example, most players would love to identify a strategy that guarantees a maximal utility. As this is not usually possible, a \emph{security strategy}, that is, a strategic choice that guarantees an explicit lower bound to the utility received, is also sought. But for a fixed $(n-1)$-tuple of opponents' strategies, rational players seek a \emph{best reply}, that is a strategy $s^{\star}_i \in S_i$ that delivers a utility at least as great, if not greater, than any other strategy $s_i \in S_i$. That is \[ G(\star, \cdots, \star, s^{\star}_i, \star, \cdots, \star) \geq G(\star, \cdots, \star, s_i, \star, \cdots, \star) \hspace{.5cm} \forall s_i \in S_i \]

A \emph{Nash equilibrium (NE)} for $G$ is a strategy profile $(s_1, s_2, ..., s_n)$ such that each $s_i$ is a best reply to the $(n-1)$-tuple of opponents' strategies. Other ways of expressing this concept include the observation that no player can increase his or her payoffs by unilaterally deviating from his or her equilibrium strategy, or that at equilibrium a player's opponents are indifferent to that player's strategic choice. As an example, consider the Prisoner's Dilemma, a two player game where each player has exactly two strategies (a so-called \emph{bimatrix} game) whose payoff function is indicated by the tableau below
\begin{table}[ht]
	\centering
	  \begin{tabular}{r|r|r|}
	    &$t_1$&$t_2$\\
		  \hline
      $s_1$&$(3,3)$&$(0,5)$\\
      \hline
			$s_2$&$(5,0)$&$(1,1)$\\
			\hline
	  \end{tabular}
	\caption{Prisoner's Dilemma}
	\label{tab:PrisonerSDilemma}
\end{table}
\noindent

Here, note that for player 1 the pure strategy $s_2$ always delivers a higher outcome than the strategy $s_1$ (say $s_2$ \emph{strongly dominates} $s_1$) and for player 2 the strategy $t_2$ strongly dominates $t_1$. Hence the pair $(s_2, t_2)$ is a (unique) Nash Equilibrium. 

However, games need not have equilibria amongst the pure strategy profiles as is exemplified by the final forms of both our poker games under consideration.  We refer to the final forms of these games because in the beginning the strategy spaces for both games are quite large, but can be simplified by making certain rationality assumptions about the players.  A common such assumption is that rational players will never play a given strategy when there is another available strategy that dominates it, that is the dominating delivers a payoff as good or better (and in at least one case strictly better) than that of the dominated strategy for every profile of opponents' strategic choices.  When this fact is known to both players, dominated strategies can be safely eliminated from the players' strategic sets with no loss of information. Thus the game theoretic analysis of many games begins with the sequential elimination of dominated strategies.  Note for our example of the Prisoner's Dilemma above, this process identifies the Nash Equilibrium $(s_2, t_2)$ almost immediately.

\section{Application to our Poker Models}
This process is utilized in the classical analysis of Simplified Poker, where for computational convenience we set the antes at 15 and the bets at 10; and in the analysis of the Nash-Shapley model where again for computational convenience we set the antes at 16 and the bets at 64.   In both Simplified Poker and the Nash-Shapley poker model the elimination process terminates when each player has but two strategies left to choose from.

What do these strategies look like ?  They are in fact functions, with domain the set of hands (in our case just individual cards) the players could possibly be dealt and range an n-tuple of players actions, a coordinate for each opportunity in the game where a player can act and an assignment to each of these opportunities of the specific action the player takes.

In Simplified Poker each player has at most one opportunity to act.  The possible actions for player 1 are to either pass, moving the game to a showdown, or to bet.  If player 1 bets, then the actions available to player 2 are to either pass, thereby awarding the pot to player 1 or to call, that is, also placing a bet in the pot and moving the game to a showdown.  Of course the hand (in this case the card) a player is dealt can affect a player's choice of action.  Thus a pure strategy for player 1 is a function with domain {$H$, $L$} and range {Pass, Bet}.  There are four such functions and the two that survive the process of eliminating of dominated strategies are playing directly, that is passing when holding an $L$ and betting when holding an $H$ (denoted in the tables as $s_1$), and running a possible bluff, i.e. betting holding an $L$ or an $H$ (denoted in the tables as $s_2$).  For player 2, initially there are also four pure strategies and after eliminating dominated strategies, her surviving strategies are either playing directly, that is, passing when holding an $L$ and calling when holding an $H$ (denoted in the tables as $t_1$) or calling a possible bluff, that is calling when holding either an $L$ or an $H$ (denoted in the tables as $t_2$). The bimatrix game arising from this particular form of Simplified Poker has payoff function given by

\begin{table}[h]
	\centering
	  \begin{tabular}{r|r|r|}
	    &$t_1$&$t_2$\\
		  \hline
      $s_1$&$(0,0)$&$(5/2,-5/2)$\\
      \hline
			$s_2$&$((5/4,-5/4)$&$(0,0)$\\
			\hline
	  \end{tabular}
	\caption{Simplified Poker}
	\label{tab:SimplifiedPoker}
\end{table}

Turning to the Nash-Shapley poker model there are now three players, and as before at every opportunity to act each player has the choice of exactly two actions, either to Pass or to Bet and as each player executes their choice of action as the play of the game proceeds, an \emph{action sequence} of the play is defined. As usual, a Pass after a Bet earlier in the action sequence constitutes folding and hence the loss of any claim to any part of the pot, and a Bet following an earlier one in the action sequence constitutes a call. But in this model a Pass by an early position player when no Bet has been made in the action sequence up to that point does not necessarily mean that that player cannot later call a bet by subsequent player.  Indeed, if player 1 elects to Pass at his first opportunity and player 2 or 3 (or both) subsequently bets, the action will return to player 1 who will then be faced with the choice of either Passing (i.e. folding) or Betting (i.e. calling) and similarly for player 2. If in the sequence of action only a single player has Bet and the others Pass, then that player is awarded the pot; and if multiple players have bet, a showdown determines the winner or winners as described above for Simplified Poker. But slightly different from that game, the action sequence in the Nash-Shapley model consisting of a Pass by all three players does not force a showdown, rather each player has their ante returned, and thus receives a payoff of 0.

In the Nash-Shapley model then we find there are 13 possible action sequences $BBB$, $BBP$, $BPB$, $BPP$, $PBBB$, $PBBP$, $PBPB$, $PBPP$, $PPBBB$, $PPBBP$, $PPBPB$, $PPBPP$, $PPP$.  Moreover, each player has exactly four possible opportunities to act, depending on the action sequence up to that point in the game, and so a pure strategy for each player must give a 4-tuple of actions for each possible card dealt.  There are 16 such 4-tuples, and hence each player has $16^2 = 256$ pure strategies to choose from.  As before we assume that our players are rational in the sense that they care only about money, more is better than less, and that all dollars are of equal utility to them.  Then we proceed to eliminate dominated strategies just as before, for example, note that any strategy that calls for a player to fold when holding $H$ is strongly dominated.  This process terminates with each player having exactly two strategies remaining, one calling for direct play, the other calling for deceptive play in the correct circumstance.  These strategies are indicated in tables 3, 4, and 5 below where we consider the strategy spaces for players 1, 2, and 3 as $S = \left\{s_1, s_2\right\}, T =  \left\{t_1, t_2\right\}$ and $U =  \left\{u_1, u_2\right\}$ respectively.
\begin{table}
 \begin{minipage}{2.5in}
  \begin{tabular}{|r|r|r|r|r|}
	  	  \hline
	    &$--$&PBB&PBP&PPB\\
		  \hline
      H&$B$&$B$&$B$&$B$\\
      \hline
			L&$P$&$P$&$P$&$P$\\
			\hline
	  \end{tabular}\\ \\
	    \centering 
	  $s_1$ Play directly
 \end{minipage}
 \begin{minipage}{2.5in}
  \begin{tabular}{|r|r|r|r|r|}
	  	  \hline
	    &$--$&PBB&PBP&PPB\\
		  \hline
      H&$P$&$B$&$B$&$B$\\
      \hline
			L&$P$&$P$&$P$&$P$\\
			\hline
	  \end{tabular}\\ \\
	    \centering 
	  $s_2$ Slow play a winner
 \end{minipage}
 \caption{\small{Player 1.}}
\end{table}
\begin{table}
 \begin{minipage}{2.5in}
  \begin{tabular}{|r|r|r|r|r|}
	  	  \hline
	    &B&P&PPBB&PPBP\\
		  \hline
      H&$B$&$B$&$B$&$B$\\
      \hline
			L&$P$&$P$&$P$&$P$\\
			\hline
	  \end{tabular}\\ \\
	    \centering 
	  $t_1$ Play directly
 \end{minipage}
 \begin{minipage}{2.5in}
  \begin{tabular}{|r|r|r|r|r|}
	  	  \hline
	    &B&P&PPBB&PPBP\\
		  \hline
      H&$B$&$P$&$B$&$B$\\
      \hline
			L&$P$&$P$&$P$&$P$\\
			\hline
	  \end{tabular}\\ \\
	    \centering 
	  $t_2$ Slow play a winner
 \end{minipage}
 \caption{\small{Player 2.}}
\end{table}
\begin{table}[t]
\begin{minipage}{2.5in}
  \begin{tabular}{|r|r|r|r|r|}
	  	  \hline
	    &BB&BP&PB&PP\\
		  \hline
      H&$B$&$B$&$B$&$B$\\
      \hline
			L&$P$&$P$&$P$&$P$\\
			\hline
	  \end{tabular}\\ \\
	    \centering 
	  $u_1$ Play directly
 \end{minipage}
  \begin{minipage}{2.5in}
  \begin{tabular}{|r|r|r|r|r|}
	  	  \hline
	    &BB&BP&PB&PP\\
		  \hline
      H&$B$&$B$&$B$&$B$\\
      \hline
			L&$P$&$P$&$P$&$B$\\
			\hline
	  \end{tabular}\\ \\
	    \centering 
	  $u_2$ Run a possible bluff
 \end{minipage}
\caption{\small{Player 3.}}
\end{table}
\newpage

Analyzing the results for each of the eight possible deals and eight pure strategy profiles we get the  table of expected payoffs to our players which appears in table 6.

\begin{table}
  \begin{minipage}{2.5in}
  \begin{tabular}{|c|c|c|}
	  \hline
	    &$t_1$&$t_2$\\
		  \hline
      $s_1$&$(0,0,0)$&$(2,-4,2)$\\
      \hline
	     $s_2$&$(-4,2,2)$&$(-3,-3,6)$\\
			\hline
	  \end{tabular} \\ \\
	  \centering 
	  $u_1$
  \end{minipage}
  \begin{minipage}{2.5in}
      \begin{tabular}{|r|r|r|}
	  	  \hline
	    &$t_1$&$t_2$\\
		  \hline
      $s_1$&$(-2,-2,4)$&$(-2,6,-4)$\\
      \hline
			$s_2$&$(6,-2,-4)$&$(10,10,-20)$\\
			\hline
	  \end{tabular}\\ \\
	  \centering 
	  $u_2$
  \end{minipage}
  \caption{\small{The Nash-Shapely payoff function. }}
 \end{table}
In table 6 the sub-table on the left is the payoff function when player 3 plays $u_1$, while the sub-table on the right is the payoff function when player 3 employs $u_2$.

\section{Classical Randomization and Game Extension}
Note now that neither Simplified Poker nor the Nash-Shapley Poker model has a Nash equilibrium among the pure strategy profiles.  Classical game theoretic formalism now calls upon the theorist to extend the game $G$ by enlarging the domain to the set of so-called \emph{mixed} strategies and then extending the payoff functions to this larger domain. Of course, the question of if and how a given function extends is a time-honored problem in mathematics. A careful application of the mathematics of extension suggests a formalism for extending these (and other) games via quantization.  These "quantized" games will exhibit drastic changes in game behavior as compared to the classical situation.  Thus we examine this extension process in some detail.

Consider then, for each player, the set of \emph{mixed strategies}, that is, the set of probability distributions over each player's pure strategy set $S_i$. For a given set $X$, denote the probability distributions over $X$ by $\Delta(X)$ and note that when $X$ is finite, with $k$ elements say, the set $\Delta(X)$ is just the $k-1$ dimensional simplex $\Delta^(k-1)$ over $X$, i.e., the set of real convex linear combinations of elements of $X$. Of course, we can embed $X$ into $\Delta(X)$ by considering the element $x$ as mapped to the probability distribution which assigns 1 to $x$ and 0 to everything else. Thus in the finite case, the pure strategies embed into the mixed strategies as the "corners" of the appropriate simplex.  For a given game $G$, denote this embedding of $S_i$ into $\Delta(S_i)$ by $e_i$.

Now our game $G$ is classically extended to a new, larger game $G^{mix}$, as follows. Given a profile $\left(p_1, \dots, p_n\right)$ of probability distributions over the $S_i$'s, by taking the product distribution we obtain a probability distribution over the product $\prod S_i$. Taking the push out by $G$ of this probability distribution we obtain a probability distribution over the image of $G$. By following this by the expectation operator, we obtain the {\it expected outcome of the mixed strategy profile} $\left(p_1, \dots, p_n\right)$. Assigning the expected outcome to each mixed strategy profile we obtain the extended game 
$$
G^{mix}: \prod \Delta(S_i) \rightarrow \prod \Omega_i
$$

Note $G^{mix}$ is a true extension of $G$ as $G^{mix} \circ \Pi e_i = G$; that is, we have the commutative diagram that appears in Figure 1.
\clearpage
\begin{figure}[h]
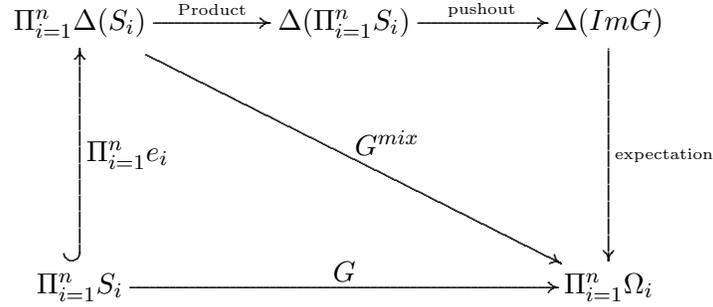

\[ \begindc{0}[50]
\obj(0,2){$\Pi_{i=1}^n \Delta(S_i)$}
\obj(2,2){$\Delta (\Pi_{i=1}^n S_i)$}
\obj(4,2){$\Delta(ImG)$}
\obj(4,0){$\Pi_{i=1}^n \Omega_i$}
\obj(0,0){$\Pi_{i=1}^n S_i$}
\mor{$\Delta (\Pi_{i=1}^n S_i)$}{$\Delta(ImG)$}{$\mbox{\tiny{pushout}}$}
\mor{$\Pi_{i=1}^n \Delta(S_i)$}{$\Delta (\Pi_{i=1}^n S_i)$}{$\mbox{\tiny{Product}}$}
\mor{$\Delta(ImG)$}{$\Pi_{i=1}^n \Omega_i$}{$\mbox{\tiny{expectation}}$}
\mor{$\Pi_{i=1}^n S_i$}{$\Pi_{i=1}^n \Delta(S_i)$}{$\Pi_{i=1}^n e_i$}[\atright, \injectionarrow]
\mor{$\Pi_{i=1}^n S_i$}{$\Pi_{i=1}^n \Omega_i$}{$G$}
\mor{$\Pi_{i=1}^n \Delta(S_i)$}{$\Pi_{i=1}^n \Omega_i$}{$G^{mix}$}
\enddc \]	
	\caption{Extension of $G$ by $G^{mix}$}
	\label{fig:ExtensionOfGByGMix}
\end{figure}

Nash's famous theorem  \cite{binmore} says that if $S_i$ are all finite, then there always exists an equilibrium in $G^{mix}$. Unfortunately, this equilibrium refered to in Nash's Theorem is called a \emph{mixed strategy equilibrium for $G$}, when it is not an equilibrium of $G$ at all, the abusive terminology confusing $G$ with its image, Im$G$. Indeed, this abusive terminology is where much confusion in quantum (and classical) game theory begins. 

For Simplified Poker above the equilibrium strategies are where player I chooses his first strategy $\frac{1}{3}$ of the time and player II chooses her second strategy $\frac{1}{3}$ of the time.  In equilibrium player one can expect a payout of 5/6, the exact amount player 2 can expect to lose.

What this says about poker is that it is better to "mix up your game" than to blindly follow a predetermined strategy.  Further, this little model shows that a large share of your expected gain from bluffing does not come from the pots you steal, but rather from the "mistaken" calls made by your opponent when she believes you may be bluffing when in fact you are not.

The Nash-Shapley Poker model shows more.  At equilibrium players 1 and 2 play deceptively (i.e their second pure strategy) with probability $p = -1 + \sqrt{7/5}$, approximately 18.3$\%$ of the time, and player three runs a possible bluff with probability $\frac{4p+8}{5p+12}$, approximately 68$\%$ of the time.  Also in equilibrium, player 3 expects to win about 80 cents and players 1 and 2 each expect to lose about 40 cents each.

This says a great deal about three handed poker.  As in Simplified Poker, it is better to "mix up your game" in the Nash-Shapley model than to blindly follow a predetermined pure strategy.  Player 3 is seen to have a definite advantage in this particular poker model due to the virtue of his \emph{position}, that is, the fact that he gets to act last.  Player 3 exploits this advantage by occasionally bluffing when his opponents have both displayed weakness by Passing, specifically by betting when holding $L$ and faced with the betting sequence $PP$.  The poker lingo refers to this as \emph{stealing} the pot and in equilibrium player 3 is attempting this about $ \frac{2}{3}$ of the time. 

But players 1 and 2 are not completely helpless.  They know that player 3 is frequently attempting to steal the pot and so each occasionally sets a trap for player 3 by \emph{slowplaying} their winner, specifically, by passing when holding $H$ in the hopes that player 3 is holding $L$ and will attempt a steal.  The poker lingo refers to this as \emph{snapping off} player 3's bluff and an early action player's slowplaying strategy as \emph{sandbagging}. Note that attempting to snap off bluffs is not without risk, as player 3 may well pass out the hand when holding $L$ and deny to players 1 or 2 their sure gain of at least half of players 3's ante in the situation where at least one of players 1 and 2  bet holding $H$ and player 3 holds $L$.  On the other hand, the rewards of successfully snapping off a bluff are great as if player 3 holding $L$ does attempt a steal in response to a slowplay by one of players 1 or 2, he loses the maximum possible of an ante plus a bet to that player.

One can ask, just how often does this happen ?  An easy application of Bayes' Rule  \cite{binmore} shows that in equilibrium the probability of at least one of players 1 and 2 is using his or her slowplaying strategy is about 28$\%$, so player 3 can expect to have his bluffs snapped off a little more than a quarter of the time.


\section{Classical Communication and Correlated Equilibria}

Before proceeding onto quantization, it is useful to place other classical game theoretical ideas such as classical mediated communication and Aumann's notion of a \emph{correlated} equilibrium into this context. One begins by observing that the function from $\prod_{i=1}^n \Delta(S_i) \rightarrow \Delta(ImG)$ is not necessarily onto. As an example consider any $2 \times 2$ game $G$. If player 1 plays his first pure strategy with probability $p$, say, and player 2 plays her second pure strategy with probability $q$, say, the resulting probability distribution over the outcomes of $G$ is given by the tableau below: 

\[
	  \begin{tabular}{r|r|r|}
	    &$t_1$&$t_2$\\
		  \hline
      $s_1$&$p(1-q)$&$pq$\\
      \hline
			$s_2$&$(1-p)(1-q)$&$(1-p)q$\\
			\hline
	  \end{tabular}
\]\\

\noindent
An easy exercise now shows that the element of $\Delta(ImG)$ represented by

\[
	  \begin{tabular}{r|r|r|}
	    &$t_1$&$t_2$\\
		  \hline
      $s_1$&$1/2$&$0$\\
      \hline
			$s_2$&$0$&$1/2$\\
			\hline
	  \end{tabular}
\]\\

\noindent
is not realizable by any choice of $p$ and $q$. One way that classical mediated communication can affect the play of games arises from this issue. Suppose players desire to have a particular probability distribution $\rho \in \Delta(ImG)$ apply to the outcomes of their play of the game $G$ and that during pre-play negotiation the players are able to hire a referee for negligible cost. For a given $\rho \in \Delta(ImG)$ the referee is meant to enforce $\rho$ over the outcomes of the game as follows. The referee secretly observes a random event with probability distribution $\rho$, thus determining a specific outcome of $G$. The referee then communicates to each player only his or her strategic choice that yields the observed outcome.

Note that the players are no longer playing the game $G$, but in fact a much larger game $G_{\rho}^{com}$ which is easily described for $2 \times 2$ games and whose generalization to games with larger strategic spaces should be clear from our description. Suppose the strategic space for each player is represented by the pair $S=\left\{A, B\right\}$. The strategic spaces for $G_{\rho}^{com}$ can be represented by the quadruple $T=\left\{A',B',C',D'\right\}$ where the strategy $C'$ represents a player always cooperating with the referee, $D'$ represents the strategy where the player always deviates from the referee's instruction (i.e. playing $B$ when he hears $A$ and vice-versa), $A'$ represents cooperating with the referee when $A$ is recommended and deviating otherwise, and $B'$ represents cooperating with the referee when $B$ is recommended and deviating otherwise.

Two other things to note here are, first, if both players choose to play $C'$, then the outcome of the new game is exactly the expected outcome of the game $G$ under $\rho$. Second, $G_{\rho}^{com}$ extends the original game $G$ as there are embeddings $f_i:\left\{A, B\right\} \rightarrow \left\{A',B', C', D'\right\}$ taking $A$ to $A'$ and $B$ to $B'$ such that $G=G_{\rho}^{com} \circ \prod_{i=1}^2 f_i$, as in the diagram in Figure 2.

\begin{figure}[h]
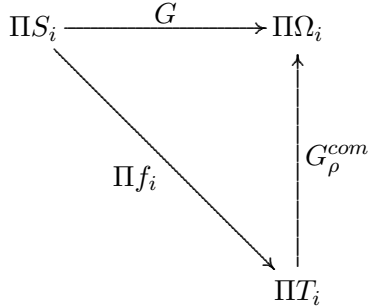

\[ \begindc{0}[50]
\obj(2,0){$\Pi T_i$}
\obj(2,2){$\Pi \Omega_i$}
\obj(0,2){$\Pi S_i$}
\mor{$\Pi S_i$}{$\Pi \Omega_i$}{$G$}
\mor{$\Pi T_i$}{$\Pi \Omega_i$}{$G_{\rho}^{com}$}[\atright, \solidarrow]
\mor{$\Pi S_i$}{$\Pi T_i$}{$\Pi f_i$}[\atright, \solidarrow]
\enddc \]	
	\caption{Extension of $G$ by $G_{\rho}^{com}$}
	\label{fig:ExtensionOfGByGCom}
\end{figure}

\noindent

Hence, classical mediated communication gives a \emph{family}, indexed by $\Delta(ImG)$, of extensions of $G$.  Following Aumann \cite{binmore}, a \emph{correlated equilibrium for} $G$ occurs whenever $(C', C')$ is a Nash equilibrium in $G_{\rho}^{com}$. That is, the players' agreement to follow the referee is \emph{self policing}, meaning that there is no gain to a player from unilateral deviating from the referee's recommendations. Note again the abusive terminology, the strategic choice for a correlated equilibrium is not a strategic choice for $G$ at all, but rather a strategic choice outside the embedded strategies for $G$ in a larger game. Of course, the use of correlated equilibrium may or may not improve the lot of the players. A classic example of correlated equilibrium improving the players' lot is given by the variant of the $2 \times 2$ game of Chicken given below
\begin{table}[h]
	\centering
	  \begin{tabular}{r|r|r|}
	    &$t_1$&$t_2$\\
		  \hline
      $s_1$&$(2,2)$&$(0,3)$\\
      \hline
			$s_2$&$(3,0)$&$(-1,-1)$\\
			\hline
	  \end{tabular}
	\caption{Chicken}
	\label{tab:Chicken}
\end{table}

An easy exercise shows that $(s_2, t_1)$ and $(s_1, t_2)$ are both pure strategy equilibria and there is a unique mixed strategy equilibrium where every player plays each of his or her pure strategies with equal probability. This mixed strategy equilibrium pays out 1 to each player. It is also easy to see that even without a referee any real convex linear combination of these three outcomes forms a self-policing agreement between the players. For example, the players could jointly observe a fair coin and agree to play the $(s_1, t_2)$ if it falls Heads and $(s_2, t_1)$ if it falls Tails. Note that the expected outcome of this agreement is $\left(\frac{3}{2}, \frac{3}{2}\right)$ which is better than the outcome $(1, 1)$ from the mixed strategy equilibrium. But even better and outside this region is the correlated equilibrium arising from the probability distribution $\frac{1}{3}(2,2)+\frac{1}{3}(0,3)+\frac{1}{3}(3,0)$ yielding the outcome $(\frac{5}{3},\frac{5}{3})$.

An example where mediated communication does not improve the lot of the players is given by Prisoner's Dilemma. One easily checks that due to the strong domination present in each player's strategy set, players always have an incentive to deviate from the referee's instruction if $\rho$ assigns a non-zero probability to any outcome other than the Nash equilibrium $(s_2, t_2)$. This domination is so strong that not even mixed strategy equilibrium that assign non-zero probability to an outcome other than the Nash equilibrium $(s_2, t_2)$ exist. A similar phenomenon occurs in the zero-sum game of Simplified Poker where any deviations from the equilibrium strategies where player I chooses his first strategy $\frac{1}{3}$ of the time and player II chooses her second strategy $\frac{1}{3}$ of the time is fully exploitable by the other player and hence an incentive to deviate from any other potential correlated equilibrium strategy.  The same holds for the Nash-Shapley Poker model.  

\section{Quantum Randomization and Game Extension}

We now wish to pass to a more general notion of randomization, that of quantum superposition and incorporate this notion into our understanding of information, communication, and games. 

Begin then with a Hilbert space $\mathcal{H}$, that is, a complex vector space equipped with an inner product. For the purpose here assume that $\mathcal{H}$ is finite dimensional, and that we have a finite set $X$ which is in one-to-one correspondence with an orthogonal basis $\mathcal{B}$ of $\mathcal{H}$. 

By a {\it quantum superposition} of $X$ (with respect to the basis $\mathcal{B}$) we mean a complex projective linear combination of elements of $X$; that is, a representative of an equivalence class of complex linear combinations where the equivalence between combinations is given by non-zero scalar multiplication. Quantum mechanics call this scalar a {\it phase}. When the context is clear as to the basis to which the set $X$ is identified, denote the set of quantum superpositions for $X$ as $QS(X)$. Of course, it is also possible to define quantum superpositions for infinite sets, but for the purpose here, one need not be so general. With care, what follows can be directly generalized to the infinite case. By the coordinatization axiom of quantum mechanics, the set $QS(X)$ corresponds to the states of a quantum system with observational basis $\mathcal{B}$.

As the underlying space of complex linear combinations is a Hilbert space, we can assign a length to each complex linear combination and, up to phase, always represent a projective linear combination by a complex linear combination of length 1. This process is called {\it normalization} and is frequently useful. 

By the measurement axiom of quantum mechanics, for each quantum superposition of $X$ we can obtain a probability distribution over $X$ by assigning to each component the ratio of the square of the length of its coefficient to the square of the length of the combination. For example, the probability distribution produced from $\alpha x + \beta y$ is just
$$
\frac{\left|\alpha\right|^2}{\left|\alpha\right|^2+\left|\beta\right|^2}x+\frac{\left|\beta\right|^2}{\left|\alpha\right|^2+\left|\beta\right|^2}y
$$
Call this function $QS(X) \rightarrow \Delta(X)$ a {\it quantum measurement with respect to $X$}, and note that geometrically quantum measurement is defined by projecting a normalized quantum superposition onto the various elements of the normalized basis $\mathcal{B}$. Denote this function by $q_X^{meas}$, or if the set $X$ and basis $\mathcal{B}$ are clear from the context, by $q^{meas}$. 

Now given a finite $n$-player game $G$, we can generalize the notion of a mixed strategy to that of a \emph{quantum strategy}.  Suppose we have a collection $Q_1, \dots, Q_n$ of non-empty sets and a \emph{protocol}, that is, a function $\cal{Q}$ $:\prod Q_i \rightarrow QS(\rm{Im}G)$. Quantum measurement $q_{\rm{Im}G}^{meas}$ then gives a probability distribution over $\rm{Im}G$. Just as in the mixed strategy case we can then form a new game $G^{\cal Q}$ by applying the expectation operator. Call the game $G^{\cal Q}$ thus defined to be the {\it quantization of $G$ by the protocol $\mathcal{Q}$}. Call the $Q_i$'s sets of {\it pure quantum strategies} for $G^{\cal Q}$. Moreover, if there exist embeddings $e'_i:S_i \rightarrow Q_i$ such that $G^{\cal Q} \circ \prod e'_i=G$, call $G^{\cal Q}$ a {\it proper} quantization of $G$. If there exist embeddings $e''_i:\Delta(S_i) \rightarrow Q_i$ such that $G^{\cal Q} \circ \prod e''_i=G^{mix}$, call $G^{\cal Q}$ a {\it complete} quantization of $G$. These definitions are summed up in the commutative diagram in Figure 3. Note that for proper quantizations the original game is obtained by restricting the quantization to the image of $\prod e'_i$. For general extensions, the game theory literature refers to this as \emph{recovering} the game $G$. 

\begin{figure}[h]
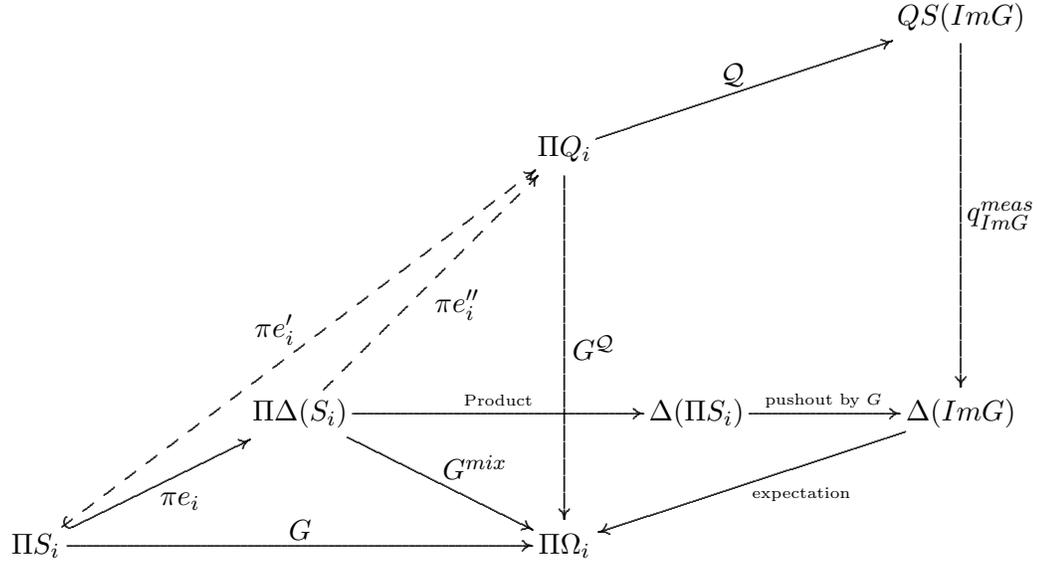

	\[ \begindc{0}[50]
\obj(4,3){$\Pi Q_i$}
\obj(7,4){$QS(ImG)$}
\obj(7,1){$\Delta(ImG)$}
\obj(5,1){$\Delta (\Pi S_i)$}
\obj(2,1){$\Pi \Delta(S_i)$}
\obj(4,0){$\Pi \Omega_i$}
\obj(0,0){$\Pi S_i$}
\mor{$\Pi Q_i$}{$QS(ImG)$}{$\cal Q$}
\mor{$\Pi Q_i$}{$\Pi \Omega_i$}{$G^{\cal Q}$}[\atleft, \solidarrow]
\mor{$QS(ImG)$}{$\Delta(ImG)$}{$q_{ImG}^{meas}$}
\mor{$\Delta(ImG)$}{$\Pi \Omega_i$}{$\mbox{\tiny{expectation}}$}
\mor{$\Pi S_i$}{$\Pi \Omega_i$}{$G$}
\mor{$\Pi \Delta(S_i)$}{$\Pi Q_i$}{$\pi e^{\prime \prime}_i$}[\atright, \dasharrow]
\mor{$\Pi S_i$}{$\Pi Q_i$}{$\pi e^{\prime}_i$}[\atleft, \dasharrow]
\mor{$\Pi S_i$}{$\Pi \Delta(S_i)$}{$\pi e_i$}[\atright, \injectionarrow]
\mor{$\Pi \Delta(S_i)$}{$\Delta (\Pi S_i)$}{$\mbox{\tiny{Product}}$}
\mor{$\Delta (\Pi S_i)$}{$\Delta(ImG)$}{$\mbox{\tiny{pushout by $G$}}$}
\mor{$\Pi \Delta(S_i)$}{$\Pi \Omega_i$}{$G^{mix}$}
\enddc \]
	\caption{A Quantization Formalism}
	\label{fig:ACompleteQuantizationOfG}
\end{figure}

Note that the definitions of $G^{mix}$ and $G^{\cal Q}$ show that a complete quantization is proper. Furthermore, note that finding a proper quantization of a game $G$ is now just a mathematically typical problem of extending a function. It is also worth noting here that nothing prohibits us from having a quantized game $G^{\cal Q}$ play the role of $G$ in the classical situation and by considering the probability distributions over the $Q_i$, creating a yet larger game $G^{m \cal{Q}}$, the {\it mixed quantization of G with respect to the protocol $\cal{Q}$}. For a proper quantization of $G$, $G^{m \cal{Q}}$ is an even larger extension of $G$. The game $G^{m \cal{Q}}$ is described in the commutative diagram of Figure \ref{fig:ExtensionOfGByGQm}.

\begin{figure}[h]
\[ \begindc{0}[50]
\obj(0,2){$\Pi_{i=1}^n \Delta(Q_i)$}
\obj(2,2){$\Delta (\Pi_{i=1}^n Q_i)$}
\obj(5,2){$\Delta(ImG^{Q})$}
\obj(5,0){$\Pi_{i=1}^n \Omega_i$}
\obj(0,0){$\Pi_{i=1}^n Q_i$}
\mor{$\Delta (\Pi_{i=1}^n Q_i)$}{$\Delta(ImG^{Q})$}{$\mbox{\tiny{pushout by $G^Q$}}$}
\mor{$\Pi_{i=1}^n \Delta(Q_i)$}{$\Delta (\Pi_{i=1}^n Q_i)$}{$\mbox{\tiny{Product}}$}
\mor{$\Delta(ImG^{Q})$}{$\Pi_{i=1}^n \Omega_i$}{$\mbox{\tiny{expectation}}$}
\mor{$\Pi_{i=1}^n Q_i$}{$\Pi_{i=1}^n \Delta(Q_i)$}{$\pi \tilde{e}_i$}[\atright, \injectionarrow]
\mor{$\Pi_{i=1}^n Q_i$}{$\Pi_{i=1}^n \Omega_i$}{$G^{\cal Q}$}
\mor{$\Pi_{i=1}^n \Delta(Q_i)$}{$\Pi_{i=1}^n \Omega_i$}{$G^{m \cal{Q}}$}
\enddc \]	
	\caption{Extension of $G$ by $G^{m \cal{Q}}$}
	\label{fig:ExtensionOfGByGQm}
\end{figure}

Note that the quantum strategy sets $Q_i$ need not consist of quantum superpositions, although in many quantization protocols they do, see for example \cite{Eisert, Landsburg}. Indeed, protocols with classical inputs yielding quantum superpositions of the outcomes of certain games have already been posited \cite{Dahl, Iqbal, Van Enk}. These and some other specific protocols are discussed in the context of the formalism of quantized analogues of classical mixed strategies given above and also in the context of a formalism for quantized analogues of classical behavioral strategies in \cite{Bleiler}.

Many protocols depend on an initial state of the quantum system in question.  When this system consists of more than a single quantum object, the individual objects may possess joint states that correlate the individual quantum states of the various objects.  This phenomenon is called {\it entanglement} and is underlying phenomenon providing the power to many quantum algorithms \cite{DJ} and the improvements in payoffs for players of quantum games \cite{Bleiler}.  

\section{Accessing Quantum Strategies-Quantum Mediated Communication and the EWL Protocol}

As discussed above, in classical mediated communication, players have a referee mediate their game and the communication of their strategic choices. When our players have but two classical pure strategies to choose from, the communication of each players strategic choices is implemented by the sending of bits to the players, put into an initial state by the referee. Presumably players then send back their individual bits in the other state (Flipped) or in the original state (Un-Flipped) to indicate the choice of their second or first classical pure strategy respectively. The returned bits are examined by the referee, who then makes the appropriate payoffs. 

When the communication between the referee and the players is over quantum channels, Eisert, Wilkens and Lewenstein \cite{Eisert} have proposed families of quantization protocols to give players access not just to mere probabilistic mixtures of their strategic choices, but also access to quantum superpositions and even probabilistic mixtures of quantum superpositions of their original strategic choices. When there are two strategic choices for each player in the basic game, players and the referee communicate over quantum channels via {\it qubits}, a two pure state quantum system with a fixed observational basis.  This observational basis is given in the so-called {\it Dirac notation} by $\left|0\right\rangle$  and $\left|1\right\rangle$. This basis also induces an observational basis of the space of the joint states of the player's qubits, denoted for the two player case in the Dirac notation by $\left|0\right\rangle \otimes \left|0\right\rangle = \left|00\right\rangle $,  $\left|0\right\rangle \otimes \left|1\right\rangle = \left|01\right\rangle$,  $\left|1\right\rangle \otimes \left|0\right\rangle = \left|10\right\rangle$, and $\left|1\right\rangle \otimes \left|1\right\rangle = \left|11\right\rangle$.   Similarly in the three player case, we have the induced observational basis for the three way joint states written in the Dirac notation as  $\left|000\right\rangle$, $\left|001\right\rangle$, $\left|010\right\rangle$, $\left|011\right\rangle$, $\left|100\right\rangle$, $\left|101\right\rangle$, $\left|110\right\rangle$, $\left|111\right\rangle$.

Each EWL protocol depends on an initial joint state of the players' qubits prepared by the referee. General actions on a qubit are represented by the elements of the special unitary group $SU(2)$ and for our game, the strategic choice represented in classical mediated communication by No Flip is now represented by the identity transformation in $SU(2)$, the strategic choice represented in classical mediated communication by Flip is now represented by an element of Lie group $SU(2)$ which interchanges the pure states of the original observational basis but also maps the initial joint state prepared by the referee under the various profiles of actions of No Flip and Flip to a set of mutually orthogonal joint states.  This set of mutually orthogonal joint states forms an alternative observational basis of the joint state space that the referee will use to determine the outcomes, and hence the payoffs for each play of the game.  In the two player case these basis elements of the joint state space are written in the Dirac notation by $\left|NN\right\rangle$, $\left|NF\right\rangle$, $\left|FN\right\rangle$, and $\left|FF\right\rangle$, and in the three player case by $\left|NNN\right\rangle$, $\left|NNF\right\rangle$, $\left|NFN\right\rangle$, $\left|NFF\right\rangle$, $\left|FNN\right\rangle$, $\left|FNF\right\rangle$, $\left|FFN\right\rangle$, and $\left|FFF\right\rangle$.

However, upon receipt of their individual qubits, players may choose not just from the matrices representing No Flip and Flip but rather from any element of the Lie group $SU(2)$ as one of their pure quantum strategies (i.e. the $Q_i$'s in the formalism above) or even probabilistic combinations thereof (i.e. the $\Delta Q_i$'s in the formalism) as their strategic choice and act on their repective qubit accordingly before returning their qubit to the referee.  Note that in practice the elements of $SU(2)$ here represent quantum superpositions of the player's original two strategies, and the mixed quantum strategies are regular probalistic combinations of these superpositions.  Note that the players have a vastly broader strategic selection in $G^{\cal{Q}}$ and $G^{m \cal{Q}}$, even when compared to the already enlarged mixed strategy game $G^{mix}$.

The payoffs to each player of each quantum or mixed quantum strategy profile are computed by the referee by observing the final joint state of the player's qubits with respect to the alternative observational basis of the joint state space described above and the referee then makes the appropriate payoffs. Per our formalism above, this procedure describes for each initial state $I$, a protocol ${\cal{Q}}_I$ and a quantized and mixed quantized game $G^{{\cal{Q}}_I}$ and $G^{{m\cal{Q}}_I}$. 

For two players, if the initial joint state prepared by the referee is given in the Dirac notation by $ \left|00\right\rangle$, then the corresponding EWL protocol is not only a complete quantization but is in fact equivalent to the classical game $G^{mix}$. This is also true in the three player case with initial joint state $ \left|000\right\rangle$. But when the initial state is given by the maximally entangled state $I = \left|00\right\rangle +\left|11\right\rangle$ in the two player case and in the case of three players by $I = \left|000\right\rangle +\left|111\right\rangle$, the corresponding EWL protocol still induces a complete quantization of the original game, but is not equivalent to the game $G^{mix}$, and in contrast to the mixed strategy situation, the corresponding protocols set up  onto maps from the appropriate product of the strategy spaces to $\Delta\left(ImG\right)$.

\section{ Payoff Calculation and Model Applications }

The next issue to address is the actual computation of the specific probability distribution over $Im(G)$ that arises from a specific profile of players' choices of elements of $SU(2)$ or, even worse, a profile of players' choices of probability distributions over $SU(2)$.  For this task it is useful to employ the quaternions, which are a non-commutative, four dimensional, normed real division algebra with canonical basis consisting of the real number $1$ and units $i$, $j$, and $k$.  These fundamental units satisfy the so-called \emph{Hamiliton relation} $ i^2 = j^2 = k^2 = ijk = -1$.  This means that each quaternion $q$ can be expressed as a real linear combination $ q = a + bi + cj + dk$ and two such are added or multiplied polynomially, subject to Hamilton's relation above.  Each quaternion $q$ as above possesses a quaternionic conjugate $q^*$ with $q^* = a - bi - cj - dk$.  The real valued multiplicative \emph{norm}  (or \emph{length} on the quaternions is defined by the formula $|q|^2 = q^*q = a^2 + b^2 + c^2 + d^2$ and all non-zero quaternions $q$ posess a non-zero inverse $q^{-1} = \frac{q^*}{|q|}$.  The \emph{unit} quaternions are those with length $1$.

For two player games, by appropriately identifying each players' pure quantum strategies with unit quaternions,  S. Landsburg \cite{Landsburg} showed in 2005 that the probability distribution over the outcomes of $G$ arising  from the profile $(p,q)$ of quantum strategies in the game  $G^{{\cal{Q}}_I}$ can be computed directly from the unit quaternion $pq$ by merely squaring the real length of each of its canonical components. This description additionally provided the computational capability to compute the expected payoffs in the game $G^{{m\cal{Q}}_I}$ by integrating over the $3$-sphere the expression $pq$ with respect to the probability distributions over the unit quaternions that form a strategy profile in $G^{{m\cal{Q}}_I}$.  This computational capability allowed Lansburg to completely determine the potential Nash equilibria of the games $G^{{\cal{Q}}_I}$ and $G^{{m\cal{Q}}_I}$, that is, the game $G$ played under the maximally entangled EWL protocol described above.  In particular, for zero sum games like Simplified Poker, Landsburg shows that there are no Nash equilibria among the pure quantum strategies, (i.e. in $G^{{\cal{Q}}_I}$) and a unique Nash equilibrium in $G^{{m\cal{Q}}_I}$ in which each player uses the uniform probability distribution (or a discrete equivalent) over his or her choice of pure quantum strategies.  The resulting probability distribution over the payoffs of $G$ is now again the uniform distribution, assigning an equal probability to each of the four outcomes of $G$ and returning as payoffs to the players the average of the original four payoffs of the Game $G$.  It is also worth noting that  the equilibrium strategies given by the uniform distribution over the pure quantum strategies are more than just best replies to each other, they are in fact security strategies against which each player's opponent has no recourse.

It is here that a fundamental question arises; that is, is this or any other Nash equilibria in such a quantized game truly new? That is, is the probability distribution that arises from an equilibrium pair in the quantized version of game $G$ different from that arising from a classical correlated equilibrium for $G$?  Note that for the Prisoner's Dilemma, this distribution does {\it not} arise from a classical correlated equilibrium as it assigns a non-zero probability to each of the classical non-equilibrium payoffs, and so does not correspond to any classical correlated equilium for this game, yet delivering a payoff $(2.5)$ to the players, that is clearly superior to the payoff $(1)$ of the classical pure strategy equilibrium. 

An even more remarkable result holds true for the maximally entangled EWL quantization of the zero-sum game of Simplified Poker, where the payout to player 1 of $\frac{15}{16}$ for the uniformly mixed quantum strategy is superior to the classical mixed strategy equilibrium payoff of $\frac{5}{6}$ for player 1, yet is still a security strategy for player 1 against which player 2 has no recourse. In particular, this shows that Player 1 can actually do better playing entangled poker over the quantum internet than playing standard  poker in either the current on-line environment or in "real life".  Player 1's advantage over player 2 in this game has actually been enhanced by quantization.

Similar, yet more dramatic changes occur in the Nash-Shapley poker model when it is played under the maximally entangled EWL protocol described above.  This follows from an extension of Landsburg's ideas by Ahmed, Bleiler, and Khan \cite{Ahmed} who,  as Landsburg, identify the players' individual quantum strategies with  copies of the unit quanternions, but then further embed these quaternionic spaces within the unit octonions.   The octonions are a non-associative, non-commutative, eight dimensional, normed real division algebra with canonical basis consisting of the real number $1$ and units $i_1$, $i_2$, $i_3$, $i_4$, $i_5$, $i_6$, $i_7$.  The elements of the octonions are expressed in the form 
\begin{equation}
\label{eq_3}
a_0 +\sum\limits_{j=1}^7 {a_j } i_j 
\end{equation}
where $a_0$ and the $a_j$ are elements of the real numbers $\mathbb{R}$.  The $i_j $'s have the property that $i_j 
^2=-1$. As before addition and multiplication are performed polynomially, subject to certain relations.  In particular, various triples of the seven $i_j $'s (along with the real number 1) can be used to 
form seven canonical copies of the quaternions $\mathbb{H}$ embedded within the octonions $\mathbb{O}$ 
as various triple products of the $i_1 \ldots i_7 $ are also equal to $-1$. A 
mnemonic for which of these triple products are in fact equal to $-1$ is given 
by an edge oriented Fano plane illustrated in Figure 5.
\begin{figure}
\begin{center}
\includegraphics[scale=0.80]{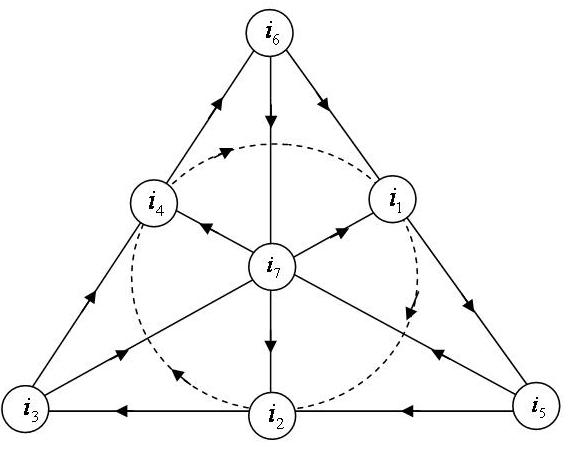}
\caption{An edge oriented Fano plane.}
\end{center}
\end{figure}

This Fano plane indicates how certain octonionic triple products work. In particular, when the $i_j ,i_k 
,i_l $ are cyclically ordered as in the lines of the edge oriented Fano plane of Figure 4, then $i_j ^2=i_k ^2=i_l ^2=i_j i_k i_l =-1$. This shows that in general $i_j 
i_k =i_l =-i_k i_j $. 

Each octonion $o$ as above possesses a octonionic conjugate $o^*$ given by the formula  
\begin{equation}
\label{eq3}
a_0 -\sum\limits_{j=1}^7 {a_j } i_j 
\end{equation}
 The real valued multiplicative \emph{norm}  (or \emph{length}) on the octonions is defined by the formula $|o|^2 = o^*o$ and all non-zero octonions $o$ posess a non-zero inverse $o^{-1} = \frac{o^*}{|o|}$.  As before the \emph{unit} octonions are those with length $1$.

From the octonionic representatives $p, q, r$ of the players strategic choices, Ahmed, Bleiler, and Khan \cite{Ahmed} then give an algebraic  octonionic expression from which the probability distribution over the outcomes of $G$ arising  from the profile of $(p,q, r)$ can be computed directly from the particular unit octonions  arising from their expression by again just squaring the real length of each of its eight canonical components.  As in the two player case, this octonionization of the payoff probabilities is used to examine the existence of Nash equilibria in the quantized versions $G^{{\cal{Q}}_I}$ and $G^{{m\cal{Q}}_I}$ of the model.  It follows for zero sum games such as the Nash-Shapley Poker model, that again there are no Nash equilibria among the pure quantum strategies, (i.e. in $G^{{\cal{Q}}_I}$) and that there exists a Nash equilibrium in $G^{{m\cal{Q}}_I}$ in which each player uses the uniform probability distribution (or a discrete equivalent) over his or her choice of pure quantum strategies.  It is also worth noting that  again the equilibrium strategies given by the uniform distribution over the pure quantum strategies are more than just best replies to each profile of opponents strategies, they are in fact security strategies against which players' opponents have no recourse.

For each player, in equilibrium the resulting probability distribution over the payoffs is again the uniform distribution, assigning an equal probability to each of the eight outcomes of the Nash-Shapley model, and thus returning as payoffs to the players the average of the original eight payoffs.  So in equilibrium players 1 and 2 can each expect to win 87.5 cents and player 3 can expect to lose $\$1.75 $.
This is an enormous change from the classical equilibrium payouts of an approximate 40 cent loss to players 1 and 2 and approximate 80 cent win for player 3.  The winners and losers of the game have been interchanged by quantization, completely destroying player 3's positional advantage, and players 1 and 2 can expect to win even more in the quantized version of the Nash-Shapley Poker model than player 3 expected to win in the classical version.  Once again we see the early position players doing far better playing entangled poker over the quantum internet than by playing standard  poker in either the current on-line environment or in "real life".


\begin{thebibliography}{99}
\bibitem{Ahmed}
A.Ahmed, S.Bleiler, and F. Khan,
{\em Three player, two strategy maximally entangled quantum games}, to appear in Proceedings, 9th International Conference in Pure Mathematics, Pakistan Mathematical Society, Islamabad, Pakistan.


\bibitem{binmore}
K.~Binmore,
{\it Fun and Games: A Text on Game Theory}, D.C. Heath (October 1991).

\bibitem{Bleiler} 
	S. A. Bleiler,
{\em A Formalism for Quantum Games}, preprint, Portland State University.


\bibitem{Borel}
E. Borel, {\em Traite du Calcul des Probabilities et ses Applications Volume IV}, Fascicule 2, Applications aux jeux des hazard, Gautier-Villars, Paris

\bibitem{Dahl}
G. Dahl, S. Landsburg,
{\em Quantum Strategies in Non cooperative Games}, University of Rochester, preprint.

\bibitem{DJ}
D. Deutsch, R. Jozsa,
{\em Rapid Solutions of Problems by Quantum Computation}, Proceedings of the Royal Society of London A 439: 553.
	
	\bibitem{Eisert}
J. Eisert, M. Wilkens, M. Lewenstein
{\em Quantum Games and Quantum Strategies}, Physical Review Letters, 
Vol. 83, Number 15, October 11, 1999.

 
 \bibitem{Iqbal}
  Azhar Iqbal, Taksu Cheon,
 {\em Constructing quantum games from nonfactorizable joint probabilities}, Phys. Rev. E 76, 061122 (2007).


 \bibitem{Landsburg}
S. Landsburg,
{\em Nash Equilibria in Quantum Games,} University of Rochester - Center for Economic Research 
(RCER) Working Paper Number 524.

	
	\bibitem{Meyer}
	  D. A. Meyer, \emph{Quantum Strategies}, Phys. Rev. Lett. $\mathbf{82}$, 1052-1055, 1999.

\bibitem{NS}
J.P. Nash and L.S. Shapley
{\em A Simple Three Person Poker Game}, Annals of Mathematics Study 24, Princeton University Press, Princeton, N.J.
	  
	  \bibitem{Van Enk}
S. J. van Enk, R. Pike,
{\em Classical rules in quantum games}
Physical Review A 66, 024306 2002.

\bibitem{VN}
J. von Neumann and O. Morganstern
{\em The Theory of Games and Economic Behavior}, Princeton University Press (1944).

\end{thebibliography}
\end{document}